\documentclass[12pt]{article}
\usepackage{latexsym}

\def\dag{\dagger} \def\pd{\partial} \def\pp{\prime} \def\a{\alpha} \def\b{\beta} \def\dl{\delta} \def\s{\sigma}  \def\eps{\epsilon}  \def\lam{\lambda}  \def\gm{\gamma}  \def\om{\omega} \def\Om{\Omega} \def\sq{\sqrt} \def\fr{\frac} \def\half{\frac{1}{2}}

\def\hg{{\hat g}}   \def\hnb{{\hat \nabla}} \def\hDelta{{\hat \Delta}} \def\hR{{\hat R}} \def\C{{\bf C}} \def\D{{\bf D}} \def\E{{\bf E}} \def\H{{\bf H}} \def\V3{{\rm V}_3}      \def\hq{\hat{q}} \def\hp{\hat{p}} \def\prm{m^\prime}

\def\rb{{\rm b}} \def\rc{{\rm c}} \def\gh{{\rm gh}}  \def\BRST{{\rm BRST}}
\def\l:{: \!}\def\r:{\! :}

\begin{document}

\begin{titlepage}

\vspace{5mm}

\begin{center}
{\Large {\bf BRST Analysis of Physical Fields and States for 4D Quantum Gravity on $R \times S^3$}} 
\end{center}

\vspace{5mm}

\begin{center}
{\sc Ken-ji Hamada}\footnote{E-mail address: hamada@post.kek.jp; URL: http://research.kek.jp/people/hamada/}
\end{center}

\begin{center}
{\it Institute of Particle and Nuclear Studies, KEK, Tsukuba 305-0801, Japan} \\ and \\
{\it Department of Particle and Nuclear Physics, The Graduate University for Advanced Studies (Sokendai), Tsukuba 305-0801, Japan}
\end{center}

\begin{abstract}
We consider the background-free quantum gravity based on conformal gravity with the Riegert-Wess-Zumino action, which is formulated in terms of a conformal field theory. Employing the $R \times S^3$ background in practice, we construct the nilpotent BRST operator imposing diffeomorphism invariance. Physical fields and states are analyzed, which are given only by real primary scalars with a definite conformal weight. With attention to the presence of background charges, various significant properties, such as the state-operator correspondence and the norm structure, are clarified with some examples.
\end{abstract}

\end{titlepage}

\section{Introduction}
\setcounter{equation}{0}
\noindent

Conformal field theory (CFT) appears in various branches of theoretical physics. Quantum gravity is also described as a certain CFT that has conformal invariance as gauge symmetry. It is well-known that two-dimensional (2D) quantum gravity is described in terms of the CFT called the Liouville theory and the Virasoro algebra represents the background-free picture of 2D spacetime \cite{polyakov, kpz, dk, seiberg, gl, bmp}.

The four-dimensional quantum gravity we will study here \cite{riegert, am, amm92, amm97b, hs, hamada02, hh, hamada05, hamada09a, hamada09b, hamada11, hamada12} is described in terms of such a CFT, which is formulated on the basis of conformal gravity systematically incorporating the Riegert-Wess-Zumino action induced from the path integral measure, as in the case of 2D quantum gravity. The model is characterized by how the metric field decomposes into the conformal factor $e^{2\phi}$ and the traceless tensor field $h_{\mu\nu}$ \cite{hamada02, hamada09a}:
\begin{equation}
     g_{\mu\nu}=e^{2\phi}(\hg e^{th})_{\mu\nu}= e^{2\phi} \left( \hg_{\mu\nu} + t h_{\mu\nu} + \cdots \right) ,
       \label{metric decomposition}
\end{equation}
where $tr(h)=\hg^{\mu\nu}h_{\mu\nu}=0$ and $\hg_{\mu\nu}$ is the background metric. Nonperturbative effects are incorporated by treating the conformal factor $e^{2\phi}$ exactly without introducing its own coupling constant, while the traceless tensor field $h_{\mu\nu}$ is handled by the perturbation theory. Here, $t$ is the dimensionless coupling constant indicating asymptotic freedom. The model we consider is obtained at the ultraviolet (UV) limit of $t=0$, where exact conformal invariance arises as a realization of background metric independence.

In this paper, we continue the study of physical fields and states of the model developed in \cite{hh, hamada05, hamada09b, hamada11} in the context of the Becchi-Rouet-Stora-Tyutin (BRST) quantization \cite{brs, kugo, kato, fms, bmp}. The BRST symmetry we discuss here is the residual diffeomorphism symmetry left after the gauge fixing, such that the gauge degrees of freedom reduce to the 15 conformal Killing vectors $\zeta^\mu$ satisfying $\hnb_\mu \zeta_\nu + \hnb_\nu \zeta_\mu - \hg_{\mu\nu} \hnb_\lam \zeta^\lam/2 =0$ \cite{hh, hamada09b}. The BRST transformation is obtained by replacing $\zeta^\mu$ with the corresponding gauge ghost $c^\mu$ as 
\begin{eqnarray}
     \dl_{\rm B} \phi &=& c^\lam \hnb_\lam \phi + \fr{1}{4} \hnb_\lam c^\lam ,
             \nonumber \\ 
    \dl_{\rm B} h_{\mu\nu} &=& c^\lam \hnb_\lam h_{\mu\nu} 
              + \half h_{\mu\lam} \left( \hnb_\nu c^\lam - \hnb^\lam c_\nu \right)
              + \half h_{\nu\lam} \left( \hnb_\mu c^\lam - \hnb^\lam c_\mu \right) .
           \label{BRST transformation}       
\end{eqnarray}

This transformation can be regarded as a conformal transformation considering quantum gravity as a quantum field theory on the background spacetime. Due to the presence of the shift term in the first equation, the invariance under the conformal change of the background metric occurs as the gauge symmetry. Thus, the background-free nature is represented as the gauge equivalence between the metrics before and after the change.

Unlike usual CFT, this conformal invariance is imposed on the field as well as the vacuum because it is the gauge symmetry. Although the residual gauge degrees of freedom are finite, this symmetry is much stronger because the right-hand side of (\ref{BRST transformation}) is field-dependent and so the transformation mixes all modes in the field. Indeed, physical fields are given only by real primary scalars with a definite conformal weight.

The analysis of physical quantities is carried out employing the cylindrical background $R \times S^3$ in practice because it is useful to study physical states and their norm structures. The related study on the Minkowski background $M^4$ has been carried out in the previous work \cite{hamada12}. The result is consistent with this case, as is expected from the background-free nature of the model.

This paper is presented as follows. In the next section, we briefly summarize the model and the definitions of the basic objects we will use. The nilpotent BRST operator imposing diffeomorphism invariance is constructed in Section 3. Physical fields and states are studied in the context of BRST formalism, with some examples given in Sections 4 and 5. The physical significant properties, such as the state-operator correspondence and the norm structure, are clarified in Section 6. Section 7 is devoted to a conclusion and discussion.

\section{Brief Summary of The Model}
\setcounter{equation}{0}
\noindent

In order to discuss diffeomorphism symmetry at the quantum level, we have to specify the gravitational action. The action that governs the dynamics of the traceless tensor field is given by the Weyl action divided by $t^2$, $-(1/t^2) \int d^4 x \sq{-g}C_{\mu\nu\lam\s}^2$, where $C_{\mu\nu\lam\s}$ is the Weyl tensor. The action for the $\phi$ field, called the Riegert field, in the conformal factor is induced from the path integral measure. At the UV limit of $t=0$, it is given by the Riegert-Wess-Zumino action \cite{riegert}
\begin{eqnarray}
   S_{\rm RWZ} = -\fr{b_1}{(4\pi)^2} \int d^4 x \sq{-\hg} 
      \left\{  2\phi \hDelta_4 \phi + \left( \hat{G}_4 - \fr{2}{3} \hnb^2 \hR \right) \phi \right\} ,
          \label{RWZ action}
\end{eqnarray}
where $\sq{-g}\Delta_4$ is the conformally invariant fourth-order differential operator and $G_4$ is the Euler density. The quantities with the hat are defined in terms of the background metric $\hg_{\mu\nu}$. The coefficient $b_1$ is the positive-definite constant greater than 4,\footnote{ 
\label{footnote for b_1} It has been computed to be $b_1 = 769/180 + ( N_X + 11N_W/2 + 62 N_A )/360$, where $N_X$, $N_W$, and $N_A$ are numbers of scalar fields, Weyl fermions, and gauge fields, respectively \cite{cd, ddi, duff}. The first term comes from gravitational loop corrections, which is the sum of $-7/90$ and $87/20$ from the $\phi$ and $h_{\mu\nu}$ fields, respectively \cite{amm92, hs}.
} 
and thus the action is bounded from below (in Wick-rotated Euclidean space).

The Riegert-Wess-Zumino action has been quantized in \cite{am, amm92, amm97b, hs, hh, hamada05, hamada09b, hamada11, hamada12} and the quantization of the Weyl action has been carried out in the perturbative expansion by $t$ in \cite{hamada02, hamada09a}.\footnote{ 
The beta function has been computed to be $\b_t=-\b_0 t_r^3$, with $\b_0=\{ 197/30 +(N_X + 3N_W + 12 N_A )/120 \}/32\pi^2$, where the first term of $\b_0$ is the sum of $-1/15$ and $199/30$ from the $\phi$ and $h_{\mu\nu}$ fields, respectively \cite{cd, ddi, duff, amm92, hs}. It indicates the asymptotic freedom and ensures the positivity of the two-point function of the stress tensor for this combined system because its coefficient is given by $\b_0$. 
}  
It has been shown that the algebra of diffeomorphism symmetry (\ref{BRST transformation}) is closed at the quantum level in this system without the $R^2$ action \cite{hh, hamada09b, hamada12}, as indicated from the Wess-Zumino integrability condition \cite{wz, bcr, riegert}.

In this paper, the background is practically chosen to be $R\times S^3$ with unit $S^3$. The Riegert field is expanded in scalar harmonics on $S^3$ denoted by $Y_{JM}$ (\ref{scalar harmonics Y}), which is decomposed into three parts: creation mode, zero mode, and annihilation mode, with $\phi = \phi_> + \phi_0  + \phi_<$, where $\phi_0 = \phi_0^\dag$, $\phi_> = \phi_<^\dag$ and
\begin{eqnarray}
   \phi_0 &=& \fr{1}{\sq{2b_1}} \left( \hq + \eta \hp \right), 
                     \nonumber \\
   \phi_< &=& \fr{\pi}{2\sq{b_1}} \left\{ 
               \sum_{J \geq \half} \sum_M \fr{a_{JM} e^{-i2J\eta}Y_{JM}}{\sq{J(2J+1)}}
              +\sum_{J \geq 0} \sum_M \fr{b_{JM} e^{-i(2J+2)\eta}Y_{JM}}{\sq{(J+1)(2J+1)}}  \right\} .
           \label{mode expansion}
\end{eqnarray}
The commutation relations are given by $[\hat{q}, \hat{p}] = i$ and $\left[ a_{J_1 M_1},a^\dag_{J_2 M_2} \right] =  -\left[ b_{J_1 M_1},b^\dag_{J_2 M_2} \right] = \dl_{J_1 J_2}\dl_{M_1 M_2}$, where $a_{JM}$ and $b_{JM}$ are the positive-metric and negative-metric modes, respectively. The index $J (\geq 0)$ with $M$ in these modes and scalar harmonics denotes that these quantities belong to the $(J,J)$ representation of the $S^3$ isometry group $SU(2) \times SU(2)$ with the multiplicity $M=(m,m^\pp)$, where $m,m^\pp =-J, -J+1, \cdots, J$. The delta function for the multiplicity index is defined by $\dl_{M N}=\dl_{m n}\dl_{m^\pp n^\pp}$. For more details, see Appendix A in which the conventions and notations for indices and various tools on $R \times S^3$ are summarized.

The 15 conformal Killing vectors on $R\times S^3$ are denoted as $\zeta^\mu= \eta^\mu, \zeta_{MN}^\mu, \zeta_M^\mu, \zeta_M^{\mu*}$ and their concrete forms are gathered in Appendix A. Here and below, for simplicity, we use the indices $M$, $N$ without $J$ for the four-vector of $J=1/2$ which appears in the conformal Killing vectors, the corresponding generators, and also ghost modes introduced in the next section. The generator of diffeomorphism symmetry that forms the conformal algebra is given by $Q_\zeta = \int d\Om_3 \zeta^\mu \hat{T}_{\mu 0}$, where $\hat{T}_{\mu\nu}$ is the stress tensor derived from the combined system of Riegert-Wess-Zumino and Weyl actions.

The 15 generators for the Riegert sector are represented as follows \cite{amm97b, hh, hamada05, hamada09b}. The Hamiltonian is $H = {\hat p}^2/2 + b_1 + \sum_{J, M} \{ 2J a^\dag_{JM} a_{JM} -(2J+2)b^\dag_{JM} b_{JM} \}$, where the constant energy shift $b_1$ is the Casimir effect. The four generators of special conformal transformations have the form
\begin{eqnarray}
     Q_M &=& \left( \hbox{$\sq{2b_1}$}-i\hat{p} \right) a_{\half M}
                 \nonumber \\
         &&  +\sum_{J \geq 0}\sum_{M_1}\sum_{M_2} \C^{\half M}_{JM_1, J+\half M_2}
              \Bigl\{ \sq{2J(2J+2)} \eps_{M_1} a^\dag_{J-M_1} a_{J+\half M_2}
                 \nonumber \\
         &&
             -\sq{(2J+1)(2J+3)} \eps_{M_1} b^\dag_{J-M_1} b_{J+\half M_2}
             + \eps_{M_2} a^\dag_{J+\half -M_2} b_{J M_1} \Bigr\} ,
                 \label{generator of special conformal transformation}
\end{eqnarray}
where $\eps_M=(-1)^{m-m^\pp}$ and $\C$ is the $SU(2) \times SU(2)$ Clebsch-Gordan coefficient defined by (\ref{C function}). The Hermitian conjugates $Q_M^\dag$ are also the generators of special conformal transformations.\footnote{ 
$Q^\dag_M$ may be called the generator of translations.
} 
The six generators of the $SU(2) \times SU(2)$ rotation group on $S^3$ are denoted by $R_{MN}$, with the properties $R^\dag_{MN}=R_{NM}$ and $R_{MN}=-\eps_M \eps_N R_{-N-M}$, whose explicit forms are not depicted here. These 15 generators form the conformal algebra of $SO(4,2)$ as follows:\footnote{ 
Parametrizing the four-vector index $\{ (1/2, 1/2), (1/2, -1/2), (-1/2, 1/2),  (-1/2, -1/2) \}$ by $\{ 1,2,3,4 \}$, and setting $A_+=R_{31}$, $A_-=R_{31}^\dag$, $A_3=(R_{11}+R_{22})/2$, $B_+=R_{21}$, $B_-=R_{21}^\dag$ and $B_3=(R_{11}-R_{22})/2$, the last rotation algebra is written in the familiar form of the $SU(2)\times SU(2)$ algebra as $[A_+, A_-]=2A_3$, $[A_3, A_\pm ]=\pm A_\pm$, $[B_+, B_-]=2B_3$, $[B_3, B_\pm ]=\pm B_\pm$, where $A_{\pm,3}$ and $B_{\pm,3}$ commute. The generators $A_{\pm,3} (B_{\pm,3})$ act on the left(right) index of $M=(m,m^\pp)$. 
} 
\begin{eqnarray}
     \left[ Q_M, Q^\dag_N \right] &=& 2\dl_{MN} H + 2R_{MN},
           \nonumber \\
    \left[ H, Q_M \right] &=& -Q_M, \quad
    \left[ H, R_{MN} \right] = 0, 
           \nonumber  \\
    \left[ Q_M, Q_N \right] &=& 0, \quad
    \left[ Q_M, R_{N L} \right] = \dl_{M L} Q_N -\eps_N \eps_L \dl_{M -N} Q_{-L} ,
            \nonumber  \\
    \left[ R_{M N}, R_{L K} \right]
        &=& \dl_{M K} R_{L N}  -\eps_M \eps_N \dl_{-N K} R_{L -M}
            \nonumber \\
        && - \dl_{N L} R_{M K} +\eps_M \eps_N \dl_{-M L} R_{-N K} .
            \label{conformal algebra}
\end{eqnarray}

The significant property of the generator $Q_M$ (\ref{generator of special conformal transformation}) is that this generator mixes the positive-metric and negative-metric modes due to the presence of the last cross term. Consequently, both of these modes cannot be gauge-invariant alone, and therefore they themselves have no physical meaning.

The same situation holds in the case of the traceless tensor field as well. The generator has been constructed from the Weyl action in \cite{hh} and its physical properties have been investigated in \cite{hamada05, hamada09b} and are briefly summarized in Appendix B.

\section{BRST Operator}
\setcounter{equation}{0}
\noindent

The gauge ghost $c^\mu$ satisfying the conformal Killing equation $\hnb_\mu c_\nu + \hnb_\nu c_\mu - \hg_{\mu\nu} \hnb_\lam c^\lam/2 =0$ is expanded by 15 Grassmann modes $\rc$, $\rc_{MN}$, $\rc_M$, $\rc_M^\dag$ as
\begin{equation}
     c^\mu = \rc \eta^\mu + \sum_M \left( \rc_M^\dag \zeta^\mu_M + \rc_M \zeta_M^{\mu *} \right)
              + \sum_{M,N} \rc_{MN} \zeta_{MN}^\mu .
           \label{ghost field}
\end{equation}
We also introduce the antighost modes $\rb$, $\rb_{MN}$, $\rb_M$, $\rb_M^\dag$. Here, $\rc$ and $\rb$ are real operators and $\rc_{MN}$ and $\rb_{MN}$ satisfy the relations $\rc_{MN}^\dag = \rc_{NM}$, $\rc_{MN} = -\eps_M \eps_N \rc_{-N-M}$, $\rb_{MN}^\dag = \rb_{NM}$ and $\rb_{MN} = -\eps_M \eps_N \rb_{-N-M}$. The anticommutation relations among these modes are defined by
\begin{eqnarray}
   \{ \rb, \rc \} &=& 1, 
       \nonumber \\
   \{ \rb_{M N}, \rc_{L K} \} &=& \dl_{M L} \dl_{N K} 
                                -\eps_M \eps_N \dl_{-M K} \dl_{-N L} ,
       \nonumber \\
   \{ \rb^\dag_M , \rc_N \} &=& \{ \rb_M , \rc^\dag_N \} = \dl_{MN} .
         \label{anticommutation relations for ghost modes}
\end{eqnarray}
For later calculations, it is useful to know that the gauge ghost modes satisfy $\sum_M \rc_{MM}=0$ and $\sum_M \eps_M \rc_{-M} \rc_M =0$, and that the antighost modes also satisfy similar equations.

Using these gauge ghost and antighost modes, we can construct the 15 generators of conformal symmetry, which are given by \cite{amm97a, hamada05}
\begin{eqnarray}
    H^\gh &=& \sum_M \left( \rc_M^\dag \rb_M - \rc_M \rb_M^\dag \right) ,
                      \nonumber \\
    R^\gh_{MN} &=& - \rc_M \rb_N^\dag + \rc^\dag_N \rb_M 
                   + \eps_M \eps_N \left( \rc_{-N} \rb^\dag_{-M} - \rc^\dag_{-M} \rb_{-N} \right)
                       \nonumber \\
               && - \sum_L \left( \rc_{LM} \rb_{LN} - \rc_{NL} \rb_{ML} \right) ,
                   \nonumber \\
    Q_M^\gh &=& -2 \rc_M \rb - \rc \rb_M - \sum_L \left( 2 \rc_{LM} \rb_L  + \rc_L \rb_{ML} \right),
                      \nonumber \\
    Q_M^{\gh\dag} &=& 2 \rc_M^\dag \rb + \rc \rb_M^\dag + \sum_L \left( 2 \rc_{ML} \rb_L^\dag + \rc_L^\dag \rb_{LM} \right).
             \label{generators for gauge ghosts}
\end{eqnarray}
These generators satisfy the same conformal algebra as (\ref{conformal algebra}). In the following, we write the full generators of the conformal algebra including the gauge ghost part as
\begin{eqnarray}
    {\cal H} &=& H + H^\gh, \qquad  {\cal R}_{MN} = R_{MN} + R^\gh_{MN} ,
           \nonumber \\
    {\cal Q}_M &=& Q_M + Q_M^\gh, \qquad  {\cal Q}^\dag_M = Q^\dag_M + Q_M^{\gh\dag} .
          \label{full generator}
\end{eqnarray}

The BRST operator generating the diffeomorphism (\ref{BRST transformation}) is now given by
\begin{eqnarray}
   Q_\BRST &=& \rc H + \sum_M \left( \rc_M^\dag Q_M + \rc_M Q_M^\dag \right) + \sum_{M,N} \rc_{MN}R_{MN} 
                     \nonumber \\
           && + \half \rc H^\gh + \half \sum_M \left( \rc^\dag_M Q_M^\gh 
                              + \rc_M Q_M^{\gh \dag} \right) + \half \sum_{M,N} \rc_{MN} R_{MN}^\gh ,
            \label{BRST operator}
\end{eqnarray}
which satisfies the Hermitian condition $Q_\BRST^\dag =Q_\BRST$. It can be written in the following form:
\begin{eqnarray}
   Q_\BRST &=& \rc {\cal H} + \sum_{M,N} \rc_{MN} {\cal R}_{MN}
               - \rb M - \sum_{M,N} \rb_{MN} Y_{MN} + \hat{Q} , 
                  \label{BRST operator 2}
\end{eqnarray}
where the full generators ${\cal H}$ and ${\cal R}_{MN}$ defined above come out. The other operators $M$, $Y_{MN}$ and $\hat{Q}$ are defined by 
\begin{eqnarray}
      M &=& 2 \sum_M \rc^\dag_M \rc_M ,
           \qquad  
      Y_{MN} = \rc^\dag_M \rc_N + \sum_L \rc_{ML} \rc_{LN} ,
            \nonumber \\
     \hat{Q} &=& \sum_M \left( \rc_M^\dag Q_M + \rc_M Q_M^\dag \right) . 
\end{eqnarray}
Using the expression (\ref{BRST operator 2}), the nilpotency of the BRST operator can be shown as
\begin{eqnarray}
    Q_\BRST^2 &=& \hat{Q}^2 - M {\cal H} -2 \sum_{M,N} \rc_M^\dag \rc_N  \left[ {\cal R}_{MN}
                  + \sum_L \left( \rc_{LM} \rb_{LN} - \rc_{NL} \rb_{ML} \right) \right]
                   \nonumber \\
              &=& \hat{Q}^2 - MH -2 \sum_{M,N}\rc_M^\dag \rc_N R_{MN} = 0 ,
\end{eqnarray}
where the conformal algebra (\ref{conformal algebra}) is used.

The anticommutation relations of the BRST operator with gauge ghost modes are
\begin{eqnarray}
    \left\{ Q_\BRST, \rc \right\} &=& -2 \sum_M \rc_M^\dag \rc_M , 
               \nonumber \\
    \left\{ Q_\BRST, \rc_{MN} \right\} &=& -\rc_M^\dag \rc_N - \eps_M \eps_N \rc_{-M} \rc_{-N}^\dag
                                                +2 \sum_L \rc_{ML} \rc_{LN} , 
               \nonumber \\
    \left\{ Q_\BRST, \rc_M \right\} &=& \rc_M \rc +2 \sum_N \rc_N \rc_{NM}  ,
               \nonumber \\
    \left\{ Q_\BRST, \rc^\dag_M \right\} &=& \rc \rc_M^\dag +2 \sum_N \rc_{MN} \rc_N^\dag 
               \label{BRST commutator for c ghost}
\end{eqnarray}
and those with antighost modes are
\begin{eqnarray}
    \left\{ Q_\BRST, \rb \right\} &=& {\cal H} , \qquad
    \left\{ Q_\BRST, \rb_{MN} \right\} = 2 {\cal R}_{MN},
               \nonumber \\
    \left\{ Q_\BRST, \rb_M \right\} &=& {\cal Q}_M , \qquad
    \left\{ Q_\BRST, \rb^\dag_M \right\} = {\cal Q}_M^\dag ,
           \label{BRST commutator}
\end{eqnarray}
where the full generators appear in the right-hand side. From (\ref{BRST commutator}), the nilpotency of the BRST operator represents $[ Q_\BRST,{\cal H} ] = [ Q_\BRST,{\cal R}_{MN} ] = [ Q_\BRST,{\cal Q}_M ] = [ Q_\BRST,{\cal Q}_M^\dag ] = 0$.

\section{Physical Fields}
\setcounter{equation}{0}
\noindent

In this section, we develop the study of physical field operators \cite{hamada11} in the context of the BRST formalism.

For each generator of the conformal algebra, the Riegert field transforms as
\begin{eqnarray}
     i \left[ H, \phi \right] &=& \pd_\eta \phi ,
            \qquad
     i \left[ R_{MN}, \phi \right] = \hnb_j \left( \zeta_{MN}^j \phi \right) ,
            \nonumber \\
     i[Q_M, \phi] &=& \zeta_M^\mu \hnb_\mu \phi +\fr{1}{4} \hnb_\mu \zeta^\mu_M  ,
            \quad
     i[Q^\dag_M, \phi] = \zeta_M^{\mu *} \hnb_\mu \phi +\fr{1}{4} \hnb_\mu \zeta^{\mu *}_M  .
        \label{comutators of conformal operator and Riegert field}
\end{eqnarray}
Here, the third equation is given by the sum of the equations
\begin{eqnarray}
     i \left[ Q_M , \phi_> \right] 
     &=& \zeta_M^\mu \hnb_\mu \phi_> + \zeta_M^0 \pd_\eta \phi_0 + \fr{1}{4} \hnb_\mu \zeta_M^\mu , 
              \nonumber \\
   i \left[ Q_M, \phi_0 + \phi_< \right] &=& \zeta_M^\mu \hnb_\mu \phi_< 
          \label{commutator with Q_M}
\end{eqnarray}
and the fourth equation is the Hermitian conjugate of the third one.

Using the BRST operator, we find that the transformation laws (\ref{comutators of conformal operator and Riegert field}) can be summarized into the single equation
\begin{equation}
     i \left[ Q_\BRST, \phi \right] = c^\mu \hnb_\mu \phi + \fr{1}{4} \hnb_\mu c^\mu .
\end{equation}
The right-hand side is the BRST transformation $\dl_{\rm B}\phi$ in (\ref{BRST transformation}). Also, the BRST transformation of the gauge ghost is given by the anticommuation relation as
\begin{eqnarray}
     i \left\{ Q_\BRST, c^\mu \right\} 
     &=& -2i \sum_M \rc_M^\dag \rc_M \eta^\mu 
         -i \sum_{M,N} \left( \rc_M^\dag \rc_N + \eps_M \eps_N \rc_{-M} \rc_{-N}^\dag \right) \zeta_{MN}^\mu
            \nonumber \\
     &&  + 2i \sum_{M,N,L} \rc_{ML} \rc_{LN} \zeta_{MN}^\mu
         + i \sum_M \left( \rc \rc_M^\dag \zeta_M^\mu + \rc_M \rc \zeta^{\mu *}_M \right)
          \nonumber \\
     &&  + 2i \sum_{M,N} \rc_{MN} \rc_N^\dag \zeta_M^\mu 
         + 2i \sum_{M,N} \rc_N \rc_{NM} \zeta_M^{\mu *}
           \nonumber \\
     &=& c^\nu \hnb_\nu c^\mu .
           \label{anti-commutator of Q_BRST and c}
\end{eqnarray}
Here, to show the second equality, we use the Grassmann nature of gauge ghosts and the product expansions of scalar harmonics (\ref{scalar harmonics product expansion}).

In the following, the BRST-invariant fields composed of only the Riegert field and the gauge ghost are studied with two examples corresponding to the cosmological constant term and the Ricci scalar curvature.

\paragraph{Cosmological constant term}
We first study the field operator given by the purely exponential function of the Riegert field. The normal ordering of such a composite operator is defined by
\begin{eqnarray}
    V_\a = \l: e^{\a \phi} \r: = \sum_{n=0}^\infty \fr{\a^n}{n!} \l: \phi^n \r: 
    = e^{\a \phi_>} e^{\a \phi_0} e^{\a \phi_<} .
          \label{definition of V}
\end{eqnarray}
The zero-mode part can be written as $e^{\a \phi_0} = e^{\hq\a/\sq{2b_1}} e^{\eta\hp\a/\sq{2b_1}} e^{-i\eta\a^2/4b_1}$. The constant $\a$, called the Riegert charge, represents a quantum correction determined by the BRST invariance condition below, which is given by a real number to reflect that $V_\a$ is a gravitational quantity.

Using the commutation relations given in (\ref{comutators of conformal operator and Riegert field}), we find that this field satisfies $i [ H, V_\a ] = \pd_\eta V_\a$ and $i [ R_{MN}, V_\a ] = \hnb_j ( \zeta_{MN}^j  V_\a )$. For the special conformal transformation, we find
\begin{eqnarray}
    i \left[ Q_M , V_\a \right] 
    = \zeta_M^\mu \hnb_\mu V_\a + \fr{h_\a}{4} \hnb_\mu \zeta_M^\mu V_\a ,
       \label{conformal transformation of V}
\end{eqnarray}
where $h_\a$ is the conformal weight of the field given by 
\begin{eqnarray}
        h_\a = \a - \fr{\a^2}{4b_1} .
        \label{definition of conformal weight h_a}
\end{eqnarray} 
Since $V_\a$ is real, the commutator between $Q_M^\dag$ and $V_\a$ is given by the right-hand side of (\ref{conformal transformation of V}) with $\zeta_M^{\mu *}$ instead of $\zeta_M^\mu$. In terms of CFT, $V_\a$ is the so-called primary scalar field with conformal weight $h_\a$. These equations are summarized into the single equation using the BRST operator as
\begin{equation}
    i \left[ Q_\BRST , V_\a \right] = c^\mu \hnb_\mu V_\a + \fr{h_\a}{4} \hnb_\mu c^\mu V_\a .   
\end{equation}

Therefore, the spacetime volume integral of $V_\a$ with definite conformal weight $h_\a =4$ commutes with the BRST operator; namely, it commutes with all generators of conformal algebra as
\begin{equation}
    i \left[ Q_\BRST , \int d\Om_4 V_\a \right] = \int d\Om_4 \hnb_\mu \left( c^\mu V_\a \right) = 0 ,
\end{equation}
where $d\Om_4 =d\eta d\Om_3$ is the spacetime volume element.

Furthermore, we can make the field locally BRST-invariant by introducing the function of the gauge ghost contracted by the totally antisymmetric $\eps$-tensor, 
\begin{equation}
     \om = \fr{1}{4!} \eps_{\mu\nu\lam\s}c^\mu c^\nu c^\lam c^\s .
         \label{ghost function}
\end{equation}
This function transforms as 
\begin{equation}
    i \left[ Q_\BRST, \om \right] = c^\mu \hnb_\mu \om = -\om \hnb_\mu c^\mu ,
\end{equation} 
where the transformation law of $c^\mu$ (\ref{anti-commutator of Q_BRST and c}) and $c^\mu \om =0$ are used. Using this commutator we can show that the product $\om V_\a$ becomes BRST-invariant without integrating over the spacetime volume as
\begin{equation}
     i \left[ Q_\BRST , \om V_\a \right] = \fr{1}{4} \left( h_\a -4 \right) \om \hnb_\mu c^\mu V_\a = 0 
\end{equation}
provided $h_\a =4$.

There are two solutions for the equation $h_\a=4$. We select the solution that approaches the canonical value $4$ in the classical limit of $b_1 \to \infty$ corresponding to the large-number limit of matter fields coupled to gravity (see footnote \ref{footnote for b_1}). The quantum cosmological constant term is thus identified to be $V_\a$ with the Riegert charge
\begin{eqnarray}
    \a = 2b_1 \left( 1 - \sq{1-\fr{4}{b_1}} \right) .
\end{eqnarray}
The constant $\a$ is real due to $b_1 >4$, as mentioned before. In the following, $\a$ takes this value.

The other solution of $h_\a=4$ is given by $4b_1-\a$ due to the duality relation $h_\a = h_{4b_1-\a}$. The operator $V_{4b_1-\a}$ does not reduce to the canonical form of the cosmological constant term at the classical limit, but two operators  $V_\a$ and $V_{4b_1-\a}$ are regarded as adjoints of one another in the presence of the background charge, as is discussed in Section 6.

\paragraph{Ricci scalar curvature}
Next, we study the field operator with derivatives. Because of the rotation invariance, the number of derivatives must be even. We here consider the real primary scalar field $W_\b$ with two derivatives that satisfies the following transformation laws: $i [ H, W_\b ] = \pd_\eta W_\b$, $i [ R_{MN}, W_\b ] = \hnb_j ( \zeta_{MN}^j  W_\b )$ and
\begin{eqnarray}
    i \left[ Q_M , W_\b \right] 
    = \zeta_M^\mu \hnb_\mu W_\b + \fr{h_\b+2}{4} \hnb_\mu \zeta_M^\mu W_\b .
      \label{commutator of Q_M and W}
\end{eqnarray} 
The equation for $Q_M^\dag$ is given by the Hermitian conjugate of (\ref{commutator of Q_M and W}). $\b$ represents the Riegert charge and $h_\b+2$ is the conformal weight of the field, where $h_\b$ is defined by (\ref{definition of conformal weight h_a}) and $2$ denotes the number of derivatives.

The conditions for the Hamiltonian and the rotation generator are rather simple, but the condition (\ref{commutator of Q_M and W}) is so strong as to determine the form of the field uniquely. We find that the following operator satisfies these conditions:\footnote{ 
This operator is slightly different from the Ricci scalar operator given in \cite{hamada11}. According to this change, we correct the discussion held there.
} 
\begin{eqnarray}
     W_\b &=& \l: e^{\b\phi} \left( \hnb^2 \phi + \fr{\b}{h_\b} \hnb_\mu \phi \hnb^\mu \phi 
                                    - \fr{h_\b}{\b} \right) \r:
                   \nonumber \\
          &=& W^1_\b + \fr{\b}{h_\b} W^2_\b - \fr{h_\b}{\b} V_\b
\end{eqnarray}
with
\begin{eqnarray}
   W^1_\b &=& \hnb^2 \phi_> V_\b + V_\b \hnb^2 \phi_<  ,
                \nonumber \\
   W^2_\b 
       &=& - \fr{1}{4}\pd_\eta \phi_0 \pd_\eta \phi_0 V_\b 
           - \half \pd_\eta \phi_0 V_\b \pd_\eta \phi_0
           - \fr{1}{4} V_\b\pd_\eta \phi_0 \pd_\eta \phi_0                
                \nonumber \\
        && - \pd_\eta \phi_0 \left( \pd_\eta \phi_> V_\b + V_\b \pd_\eta \phi_< \right) 
           - \left( \pd_\eta \phi_> V_\b + V_\b \pd_\eta \phi_< \right) \pd_\eta \phi_0
                 \nonumber \\
        && + \hnb_\mu \phi_> \hnb^\mu \phi_> V_\b +2 \hnb_\mu \phi_> V_\b \hnb^\mu \phi_< 
                + V_\b \hnb_\mu \phi_< \hnb^\mu \phi_<  ,
\end{eqnarray}
where $\hnb^2=\hnb_\mu \hnb^\mu$ and $V_\b$ has been defined by Eq. (\ref{definition of V}).

Thus, the spacetime volume integral of $W_\b$ with $h_\b =2$ commutes with all generators of the conformal algebra. It is simply expressed in terms of the BRST operator as
\begin{equation}
     \left[ Q_\BRST, \int d\Om_4 W_\b \right] =0 .
           \label{BRST invariance for int W}
\end{equation}
The quantum Ricci scalar curvature is now identified to be $W_\b$ with the Riegert charge 
\begin{eqnarray}
    \b = 2b_1 \left( 1 - \sq{1-\fr{2}{b_1}} \right) ,
\end{eqnarray}  
which is one of the solutions of $h_\b =2$. The operator $W_\b$ then reduces to the classical form of the Ricci scalar curvature $\sq{-g}R$ divided by $-6$ because $\b \to 2$ and $\b/h_\b \to 1$ in the large-$b_1$ limit.\footnote{ 
The classical form of the Ricci scalar curvature is given by $d^4 x \sq{-g}R= d\Om_4 e^{2\phi}(-6\hnb^2 \phi -6 \hnb_\mu \phi \hnb^\mu \phi + 6)$ on the $R \times S^3$ background with unit $S^3$.
} 
In the following, $\b$ is fixed to this value.

Due to the duality relation $h_\b=h_{4b_1-\b}$, another BRST-invariant operator has the form $W_{4b_1-\b}$. This operator is the adjoint of the Ricci scalar operator $W_\b$, but does not have the classical limit.

As in the case of $V_\a$, using the gauge ghost function (\ref{ghost function}), we find 
\begin{eqnarray}
     \left[ Q_\BRST, \om W_\b \right] = 0 .
        \label{BRST invariance for om W}
\end{eqnarray}
Here, note that this BRST invariance condition is stronger than the condition (\ref{BRST invariance for int W}) because the condition (\ref{BRST invariance for int W}) holds up to total divergences.

In general, as is clear from the construction, physical fields are given by primary scalar fields with conformal weight $4$, while primary tensor fields are excluded from physical fields because such fields do not become gauge-invariant under rotations and special conformal transformations due to the presence of spin terms. All of the descendant fields are excluded as well.

\section{Physical States}
\setcounter{equation}{0}
\noindent

Let us study the BRST-invariant state $|\Psi \rangle$ satisfying the condition
\begin{equation}
     Q_\BRST |\Psi \rangle  = 0 .
\end{equation}

First, we define various vacuum states. The vacuum of Fock space annihilated by the zero mode $\hat{p}$ and annihilation modes $a_{JM}$ and $b_{JM}$ is denoted by $|0\rangle$. We also introduce the conformally invariant vacuum annihilated by all generators of the conformal symmetry except gauge ghost parts, $H$, $R_{MN}$, $Q_M$, and $Q_M^\dag$, which is defined by  $| \Om \rangle = e^{-2b_1 \phi_0(0)}| 0 \rangle$, where $\phi_0(0)=\hq/\sq{2b_1}$. This vacuum and its Hermitian conjugate have the background charge $-2b_1$, respectively, and thus the vacua have the total background charge $-4b_1$.\footnote{ 
The background charge originates from the linear term in the Riegert-Wess-Zumino action (\ref{RWZ action}).
} 

The conformally invariant vacuum of the gauge ghost sector is denoted by $|0 \rangle_\gh$, which is annihilated by all generators of the gauge ghost system (\ref{generators for gauge ghosts}); namely, annihilated by all antighosts, but not annihilated by gauge ghosts. Using this, the Fock vacuum of the gauge ghost system annihilated by the annihilation modes $\rc_M$ and $\rb_M$ is given by $\prod_M \rc_M |0 \rangle_\gh$.

Since the Hamiltonian depends on neither $\rc$ and $\rc_{MN}$ nor $\rb$ and $\rb_{MN}$, the gauge ghost vacuum $\prod \rc_M|0 \rangle_\gh$ is degenerate. The degenerate partners are then given by applying $\rc$ and $\prod \rc_{MN}$ to this vacuum. The norm structure will be discussed in the next section.

For convenience, we denote the Fock vacuum state with the Riegert charge $\gm$ by
\begin{equation}
       | \gm \rangle = e^{\gm \phi_0(0)} | \Om \rangle \otimes \prod_M \rc_M |0 \rangle_\gh .
            \label{state with gamma}
\end{equation}
This state satisfies ${\cal H} |\gm \rangle = ( h_\gm -4 ) |\gm \rangle$, where $i\hat{p}|\gm \rangle =(\gm/\sq{2b_1}-\sq{2b_1} )|\gm \rangle$ is used and $-4$ comes from the gauge ghost sector.

The physical state is constructed by applying the creation modes such as $a^\dag_{JM}$, $b^\dag_{JM}$, $\rc^\dag_M$, $\rb_M^\dag$ and $\hat{p}$ to the Fock vacuum (\ref{state with gamma}), where $\hat{p}$ may be replaced by the appropriate number. Since $\{Q_\BRST,\rb \}={\cal H}$ and $\{Q_\BRST,\rb_{MN} \}=2{\cal R}_{MN}$ and the Fock vacuum is annihilated by $\rb$ and $\rb_{MN}$, we merely consider the subspace satisfying the conditions 
\begin{eqnarray}
      {\cal H} |\Psi \rangle = {\cal R}_{MN} |\Psi \rangle =0, \quad   
      \rb |\Psi \rangle = \rb_{MN} |\Psi \rangle =0 .
         \label{subspace of physical state}
\end{eqnarray}
On this subspace, from the expression of the BRST operator (\ref{BRST operator 2}), the BRST-invariant state coincides with the $\hat{Q}$-invariant state.

For the time being, we analyze physical states in the subspace (\ref{subspace of physical state}) described in the following form:
\begin{equation}
    |\Psi \rangle = {\cal A} \left( \hat{p}, a^\dag_{JM}, b^\dag_{JM}, \cdots \right) |\gm \rangle ,
      \label{restricted state of psi}
\end{equation}
where the dots denote creation modes of other fields except gauge ghosts. The operator ${\cal A}$ and the Riegert charge $\gm$ will be determined from the BRST invariance condition below. The cases in which ${\cal A}$ includes creation modes of gauge ghosts and antighosts will be discussed later.

Since $\rc_M |\Psi \rangle =0$ for the state (\ref{restricted state of psi}), the $\hat{Q}$ invariance condition is expressed as 
\begin{equation}
    \hat{Q} |\Psi \rangle = \sum_M \rc^\dag_M Q_M |\Psi \rangle =0 .
\end{equation}
Thus, together with the Hamiltonian and rotation invariance conditions in (\ref{subspace of physical state}), we reproduce the physical state conditions
\begin{eqnarray}
   (H-4) |\Psi \rangle = R_{MN} |\Psi \rangle = Q_M |\Psi \rangle =0
       \label{physical state condition}
\end{eqnarray}
studied in \cite{hh, hamada05, hamada09b}. Here, the condition for $Q_M^\dag$ is not necessary. This shows that the state $|\Psi \rangle$ is given by a primary scalar with conformal weight $4$.\footnote{ 
The primary state is, in general, defined by $H |h, \{ r \} \rangle = h |h, \{ r \} \rangle$, $R_{MN} |h, \{ r \} \rangle = \left( \Sigma_{MN} \right)_{\{ r^\pp \},\{ r \}} |h, \{ r^\pp \} \rangle$ and $Q_M |h, \{ r \} \rangle =0$, where $h$ is the conformal weight, $\{ r \}$ denotes a representation of $SU(2)\times SU(2)$, and $\Sigma_{MN}$ is the generator of spin rotations of the state. The descendant state is generated by applying $Q_M^\dag$ to the primary state $|h, \{ r \} \rangle$.
} 

The BRST invariance condition for $|\Psi \rangle$ is now equivalent to the condition that the operator ${\cal A}$ satisfies the algebra
\begin{eqnarray}
     \left[ H, {\cal A} \right] = l {\cal A}, \quad  \left[ R_{MN}, {\cal A} \right] = 0, 
     \quad  \left[ Q_M, {\cal A} \right] = 0 .
          \label{physical algebra}
\end{eqnarray}
The first condition implies that ${\cal A}$ has the conformal weight $l (\geq 0)$. By solving the Hamiltonian condition $h_\gm + l -4 = 0$ in (\ref{physical state condition}), the Riegert charge $\gm$ is determined to be
\begin{equation}
     \gm_l = 2b_1 \left( 1 - \sq{1-\fr{4-l}{b_1}} \right).
       \label{Riegert charge gamma}
\end{equation}
Here, we choose the solution where $\gm$ approaches the canonical value $4-l$ in the large-$b_1$ limit. The charges $\gm_0$ and $\gm_2$ correspond to $\a$ and $\b$ defined above, respectively.

In order to find the operator ${\cal A}$ satisfying the second and third conditions of (\ref{physical algebra}), we seek creation operators that commute with the generator $Q_M$ and then combine them in a rotation-invariant form. Since there is no creation mode that commutes with $Q_M$ for the Riegert field, we look for operators constructed in a bilinear form. Such operators have been studied previously in \cite{hamada05, hamada09b}. Using the crossing properties of the $SU(2) \times SU(2)$ Clebsch-Gordan coefficients (\ref{crossing relations}), we find that for the Riegert sector there are two types of $Q_M$-invariant creation operators with conformal weight $2L$ for integers $L \geq 1$:
\begin{eqnarray}
    S^{\dag}_{L N}
     &=& \chi(\hat{p},L) a^{\dag}_{L N}
         + \sum_{K=\half}^{L-\half} \sum_{M_1}\sum_{M_2} x(L,K)
         \C^{L N}_{L-K M_1, K M_2} a^{\dag}_{L-K M_1} a^{\dag}_{K M_2},
                \nonumber \\
    {\cal S}^{\dag}_{L-1 N}
     &=& \psi(\hat{p}) b^{\dag}_{L-1 N}
         + \sum_{K=\half}^{L-\half} \sum_{M_1}\sum_{M_2} x(L,K)
         \C^{L-1 N}_{L-K M_1, K M_2} a^{\dag}_{L-K M_1} a^{\dag}_{K M_2}
            \nonumber  \\
      && + \sum_{K=\half}^{L-1} \sum_{M_1,M_2} y(L,K)
        \C^{L-1 N}_{L-K-1 M_1, K M_2}
        b^{\dag}_{L-K-1 M_1} a^{\dag}_{K M_2} ,
           \label{building block}
\end{eqnarray}
where
\begin{eqnarray}
      x(L,K) &=& \fr{(-1)^{2K}}{\sq{(2L-2K+1)(2K+1)}}\sq{ \left( \begin{array}{c}
                                     2L \\
                                     2K
                                     \end{array} \right)
                             \left(   \begin{array}{c}
                                     2L-2 \\
                                     2K-1
                                     \end{array} \right) } 
\end{eqnarray}
and $y(L,K) = -2\sq{(2L-2K-1)(2L-2K+1)}x(L,K)$. The zero-mode operators are given by $\chi (\hat{p},L) = \sq{2} ( \sq{2b_1}-i\hat{p})/\sq{(2L-1)(2L+1)}$ and  $\psi (\hat{p}) = -\sq{2} ( \sq{2b_1}-i\hat{p})$.\footnote{ 
Here, we correct an error in the previous papers as follows: $\chi(\hat{p},L)$ is twice that given in \cite{hamada05,hamada09b}. 
} 
For any half-integer $L$ there is no such operator. The operators for the lower cases of $L$ are provided in Appendix C.

By joining these bilinear operators using the $SU(2) \times SU(2)$ Clebsch-Gordan coefficients, we can construct the basis of $Q_M$-invariant creation operators in the Riegert sector. Due to the crossing properties of the Clebsch-Gordan coefficients, any $Q_M$-invariant creation operators will be expressed in such a fundamental form. Thus, these two types of $Q_M$-invariant bilinear operators are expected to be the building blocks of physical states. Thus the physical state $|\Psi \rangle$ (\ref{restricted state of psi}) is now written in the form ${\cal A} ( S^\dag, {\cal S}^\dag, \cdots ) |\gm \rangle $, where the dots denote building blocks for other fields and all tensor indices are contracted out in an $R_{MN}$-invariant way. Since building blocks have even conformal weights, the weight $l$ for ${\cal A}$ is given by even integers, which corresponds to the number of derivatives for physical fields.

As an example, we present here the physical states corresponding to the lower cases of $l$ up to $4$. The lowest weight state is simply given by $|\a \rangle$, which corresponds to the cosmological constant term, and the second lowest state with $l=2$ is given by ${\cal S}^\dag_{00} |\b \rangle$, which corresponds to the Ricci scalar curvature, where we use the notations $\a$ and $\b$ for $\gm_0$ and $\gm_2$, respectively. For $l=4$, there are two states, $\left( {\cal S}^\dag_{00} \right)^2 |\gm_4 \rangle$ and $\sum_N \eps_N S^\dag_{1-N}S^\dag_{1N}  |\gm_4 \rangle$, where $\gm_4=0$ from (\ref{Riegert charge gamma}), which correspond to the square of the Ricci scalar and the other four derivative scalar quantities, such as the Euler density, respectively.

At $l=4$, there is another gravitational physical state. From the Weyl sector summarized in Appendix B, we find the physical state $\sum_{M,x} \eps_M c^\dag_{1(-Mx)} c^\dag_{1(Mx)} |\gm_4 \rangle$ corresponding to the square of the Weyl tensor. Here, $c^\dag_{1(Mx)}$ is the lowest creation mode of the tensor field, which is the only creation mode that commutes with $Q_M$.

For other modes in the Weyl sector, we also have to consider $Q_M$-invariant building blocks written in a bilinear form. The  purely gravitational physical state with higher conformal weight is generally given by combining building blocks for both the Riegert and Weyl sectors in a rotation-invariant way.

Finally, we discuss the cases with gauge ghost and antighost creation modes $\rc^\dag_M$ and $\rb^\dag_M$. For $l=2$, we obtain another BRST-invariant state, 
\begin{eqnarray}
     \left\{  - \left( \hbox{$\sq{2b_1}$} -i\hat{p} \right)^2 \sum_M \eps_M \rb^\dag_{-M} \rc_M^\dag  
              + \hat{h} \sum_M \eps_M a^\dag_{\half -M} a^\dag_{\half M} 
                              \right\} |\b \rangle ,
            \label{another l=2 state}
\end{eqnarray}
where $\hat{h}=\hat{p}^2/2 + b_1$. This state is, however, equivalent to the physical state given before up to the BRST trivial state.

To show this, we introduce the state
\begin{eqnarray}
    |\Upsilon \rangle = \left( \hbox{$\sq{2b_1}$} -i\hat{p} \right) 
                        \sum_M \eps_M \rb_{-M}^\dag a_{\half M}^\dag |\b \rangle 
\end{eqnarray}
satisfying the conditions ${\cal H} |\Upsilon \rangle = {\cal R}_{MN}|\Upsilon \rangle = \rb |\Upsilon \rangle = \rb_{MN} |\Upsilon \rangle =0$. Applying the BRST operator to this state, we obtain
\begin{eqnarray}
    Q_\BRST |\Upsilon \rangle 
        &=& \biggl\{ - \left( \hbox{$\sq{2b_1}$} -i\hat{p} \right)^2 \sum_M \eps_M \rb^\dag_{-M} \rc_M^\dag 
                   + 4 \left( \hbox{$\sq{2b_1}$} -i\hat{p} \right) b_{00}^\dag
                 \nonumber \\
         && \quad
              + 2\hat{h} \sum_M \eps_M a^\dag_{\half -M} a^\dag_{\half M} 
             \biggr\} |\b \rangle .
\end{eqnarray}
Thus, the state (\ref{another l=2 state}) can be written in the form 
\begin{eqnarray}
    \fr{1}{2\sq{2}} {\cal S}^\dag_{00} |\b \rangle + Q_\BRST |\Upsilon \rangle ,
\end{eqnarray}
where $\hat{h}|\b \rangle = 2|\b \rangle$ is used.

In general, it seems that the physical state depending explicitly on gauge ghosts and antighosts such as this reduces to the standard form (\ref{restricted state of psi}) up to the BRST trivial state. Thus, we only consider such a standard form throughout this paper.

\section{State-Operator Correspondences and Norm Structures}
\setcounter{equation}{0}
\noindent

In this section we discuss various significant properties such as the state-operator correspondence, the adjoint of a physical state, and the norm structure, with attention to the presence of the background charge.

Consider the physical state with the Riegert charge $\gm$ and the corresponding physical field operator $O_\gm$ satisfying the BRST invariance condition $[ Q_\BRST, \om O_\gm ]=0$. The state-operator correspondence is given by the following limit: 
\begin{eqnarray}
   \lim_{\eta \to i\infty} e^{-4i\eta}O_\gm |\Om \rangle  = |O_\gm \rangle  
\end{eqnarray}
apart from the gauge ghost sector.

For the physical fields $V_\a$ and $W_\b$, for instance, we can obtain the physical states as follows:
\begin{eqnarray}
    |V_\a \rangle = \lim_{\eta \to i\infty}e^{-4i\eta}V_\a |\Om \rangle 
    = \lim_{\eta \to i\infty} e^{i(-4+h_\a)\eta} e^{\a\phi_>}e^{\fr{\a}{\sq{2b_1}}\hat{q}} |\Om \rangle
    = e^{\a\phi_0(0)} |\Om \rangle 
\end{eqnarray}
and 
\begin{eqnarray}
   |W_\b \rangle &=& \lim_{\eta \to i\infty}e^{-4i\eta}W_\b |\Om \rangle 
           \nonumber \\
    &=& \lim_{\eta \to i\infty} e^{i(-4+h_\b)\eta} \left\{ \hnb^2 \phi_> -2i \pd_\eta \phi_> 
               + \fr{\b}{h_\b} \hnb_\mu \phi_> \hnb^\mu \phi_> \right\} 
                   e^{\b\phi_>}e^{\fr{\b}{\sq{2b_1}}\hat{q}} |\Om \rangle
          \nonumber \\
    &=& -\fr{\b}{2\sq{2}b_1}{\cal S}^\dag_{00}e^{\b\phi_0(0)} |\Om \rangle .
\end{eqnarray}
These limits exist only when $h_\a=4$ and $h_\b=2$, respectively, as is required from the physical condition.

Since the most singular term of the gauge ghost function (\ref{ghost function}) at the limit $\eta \to i\infty$ behaves as $\om \propto  e^{-4i\eta}\prod_M \rc_M$, the state-operator correspondence that includes this function is given by
\begin{eqnarray}
    \lim_{\eta \to i\infty}\om O_\gm |\Om \rangle \otimes |0 \rangle_\gh 
        \Leftrightarrow |O_\gm \rangle \otimes \prod_M \rc_M |0 \rangle_\gh .
\end{eqnarray}
The right-hand side is the physical state discussed in Section 5.

Next, we consider the adjoint of the physical state $|O_\gm \rangle \otimes \prod \rc_M|0 \rangle_\gh$. The adjoint of $|O_\gm \rangle$ is denoted by $\langle {\tilde O}_\gm|$, which is not the naive Hermitian conjugate $\langle O_\gm |$ because  in this case the Riegert charge is not conserved; namely, the zero mode does not cancel out, such that $\langle O_\gm | O_\gm \rangle$ is unnormalizable.\footnote{ 
The situation is the same as in the case of the Liouville gravity \cite{seiberg}. Unlike this case, if the Riegert charge were purely imaginary, such as $\gm=ip$, and there were no background charges, physical fields could be normalizable as $\langle O_{-ip} | O_{ip} \rangle=1$, as in the case of string theory \cite{kato, fms}. 
} 
The state $\langle {\tilde O}_\gm|$ is defined by using the other pair of the physical states derived from the duality relation $h_\gm = h_{4b_1-\gm}$.

Again, we consider the physical fields $V_\a$ and $W_\b$. The adjoints of these fields are given by
\begin{eqnarray}
     {\tilde V}_\a &=& V_{4b_1-\a},  
            \nonumber \\
     {\tilde W}_\b &=& - \fr{b_1}{4} W_{4b_1-\b} 
                    = - \fr{b_1}{4} \left( W^1_{4b_1-\b} + \fr{4b_1-\b}{h_\b} W^2_{4b_1-\b} 
                                           - \fr{h_\b}{4b_1-\b} V_{4b_1-\b} \right) 
            \nonumber \\  
            &&
\end{eqnarray}
and the out-states corresponding to these fields are
\begin{eqnarray}
     \langle {\tilde V}_\a | &=& \lim_{\eta \to -i\infty} e^{4i\eta} \langle \Om | {\tilde V}_\a 
                             = \langle \Om | e^{(4b_1-\a)\phi_0(0)} ,
               \nonumber \\
     \langle {\tilde W}_\b | 
       &=& \lim_{\eta \to -i\infty} e^{4i\eta} \langle \Om | {\tilde W}_\b 
        =  \fr{4b_1-\b}{8\sq{2}} \langle \Om | e^{(4b_1-\b)\phi_0(0)} {\cal S}_{00} .
\end{eqnarray}
They are normalized to be
\begin{eqnarray}
     \langle {\tilde V}_\a | V_\a \rangle = 1, \qquad \langle {\tilde W}_\b | W_\b \rangle = 1 .
        \label{normalization}
\end{eqnarray}
Here, $\langle \Om| e^{4b_1\phi_0(0)} |\Om \rangle =1$ is used, which comes from the Riegert charge conservation such that the charge $4b_1$ cancels the background charges in the conformally invariant in- and out-vacua.

The naive inner products between gauge ghost vacua and their Hermitian conjugates vanish as ${}_\gh\langle 0|0 \rangle_\gh = {}_\gh\langle 0|\prod \rc_M^\dag \prod \rc_M|0 \rangle_\gh=0$, which is easily confirmed by inserting the anticommutation relations $\{ \rb, \rc \} = 1$ and $\{ \rb_{M N}, \rc_{L K} \} = \dl_{M L} \dl_{N K} -\eps_M \eps_N \dl_{-M K} \dl_{-N L}$ into the relevant expressions. So, we normalize the gauge ghost sector by inserting the operator $\vartheta = i \rc \prod \rc_{MN}$ satisfying $\vartheta^\dag = \vartheta$ as 
\begin{eqnarray}
       {}_\gh\langle 0|\prod \rc_M^\dag  \vartheta \prod \rc_M |0 \rangle_\gh =1 .
\end{eqnarray} 
Thus, the adjoint of the physical state  $| O_\gm \rangle \otimes \prod \rc_M |0 \rangle_\gh$ is given by $\langle {\tilde O}_\gm | \otimes {}_\gh \langle 0| \prod \rc_M^\dag \vartheta$. In this way, we can always define the inner product of a physical state normalized to be unity.

Lastly, we mention that the result (\ref{normalization}) is consistent with the two-point correlation function calculated to be \cite{hamada11}
\begin{eqnarray}
   \langle \Om| {\tilde V}_\a(x) V_\a(0) |\Om \rangle 
   = \left( \fr{1}{L^2(\eta,\om)} \right)^4 
\end{eqnarray}
and also with the correlation function between $W_\b$ and ${\tilde W}_\b$, which will be of the same form. Here, the function $L$ is defined through the operator product $\phi(x)\phi(0) = -(1/4b_1)\times \log L^2(\eta, \om) + \l: \phi(x) \phi(0) \r:$ as
\begin{eqnarray}
   L^2(\eta,\om) = 2 \left\{ \cos \eta -\cos \fr{\om}{2} \right\} ,
\end{eqnarray}
where $-2\cos (\om/2)$ is the spatial distance and the angle $\om$ is defined in Appendix A.

\section{Conclusion and Discussion}
\setcounter{equation}{0}
\noindent

We have studied background-free quantum gravity described in terms of CFT in the context of BRST formalism. The nilpotent BRST operator generating the diffeomorphism was constructed on the $R \times S^3$ background. We used this operator to construct the BRST-invariant fields and states and studied various significant properties, such as the state-operator correspondence and the norm structure. In terms of CFT, these are given by primary scalars with definite conformal weight $4$, while primary tensors and all of their descendants are excluded.

The BRST-invariant fields always appear in pairs due to the existence of the duality in Riegert charges. The physical field was identified with the one that reduces to the classical gravitational scalar quantity in the large-$b_1$ limit corresponding to the large-number limit of matter fields coupled to gravity.

The naive inner product between the physical state and its Hermitian conjugate is unnormalizable because the Riegert charge is not conserved; namely, the zero mode does not cancel out. The adjoint of physical state is given by the other member of the BRST-invariant pair, which does not have the classical limit, and so is regarded as a quantum virtual state. With this state, the Riegert charge can be conserved and we can define the inner product normalized to be unity.

We now discuss how to define correlation functions among physical fields with the correct Riegert charge. Naively, they do not exist because the Riegert charge is not conserved as mentioned above. To define the correlation functions, we should consider (for instance) the model perturbed by the cosmological constant term, and then the constant mode of the Riegert field $\s$ should be taken into account. Carrying out the path integral over the constant mode $A=e^{\a\s}$ first (in Wick-rotated Euclidean space), we obtain the correlator in the perturbed theory $S_{\rm RWZ}+ \mu {\bar V}_\a$ as follows:     
\begin{eqnarray}
   \langle\langle {\bar O}_{\gm_{l_1}} \cdots {\bar O}_{\gm_{l_n}} \rangle\rangle 
   &=& \fr{1}{\a} \int^\infty_0 \fr{dA}{A} A^{-s} 
        \langle {\bar O}_{\gm_{l_1}} \cdots {\bar O}_{\gm_{l_n}} e^{-\mu A {\bar V}_\a} \rangle
         \nonumber \\
   &=& \mu^s \fr{\Gamma(-s)}{\a}  
          \langle {\bar O}_{\gm_{l_1}} \cdots {\bar O}_{\gm_{l_n}} \left( {\bar V}_\a \right)^s  \rangle 
\end{eqnarray}
with $s= (4b_1 - \sum_{i=1}^n \gm_{l_i})/\a$.\footnote{ 
Here, $s$ is not an integer, but a fractional number. Therefore, by regarding $s$ as an integer the correlator may be evaluated, and then $s$ may be analytically continued to the fractional number \cite{gl}. 
} 
Here, $\mu$ is the cosmological constant. The bar on the field denotes that the field is integrated over the spacetime volume and $\langle \cdots \rangle$ represents the correlator in the unperturbed theory. This correlator will exist because the Riegert charge is conserved. It indicates that the correlation function has a power-law behavior in the mass scale. Its physical implications to inflationary cosmology are discussed elsewhere \cite{hy, hhy06, hhy10}.


\appendix

\section{Basic Tools on $R \times S^3$}
\setcounter{equation}{0}
\noindent

The notations and conventions for various tools on $R\times S^3$ \cite{hh} are summarized here. The background metric is parametrized by the coordinate $x^\mu=(\eta,x^i)$ using the Euler angles $x^i=(\a,\b,\gm)$ as $d\hat{s}^2_{R\times S^3} = -d\eta^2 + \fr{1}{4} (d\a^2 +d\b^2 +d\gm^2 +2 \cos \b d\a d\gm )$, where $\a$, $\b$ and $\gm$ have the ranges $[0,2\pi]$, $[0,\pi]$, and $[0,4\pi]$, respectively. The radius of $S^3$ is taken to be unity such that $\hR=6$. The volume element on the unit $S^3$ is $d\Om_3 = \sin \b d\a d\b d\gm/8$ and the volume is given by $\V3 = \int d\Om_3 =2\pi^2$. The angle $\om$ is defined by $\cos(\om/2)= \cos(\b/2)\cos(\a/2)\cos(\gm/2)-\cos(\b/2)\sin(\a/2)\sin(\gm/2)$.

The scalar harmonics on $S^3$ are defined by 
\begin{equation}
    Y_{JM} = \sq{\fr{(2J+1)}{\V3}} D^J_{m\prm} 
     \label{scalar harmonics Y}
\end{equation} 
satisfying $\hnb^j\hnb_j Y_{J M} = -2J(2J+2) Y_{J M}$, where $D^J_{m\prm}$ is the Wigner $D$-function \cite{vmk}. It belongs to the $(J,J)$ representation of the isometry group $SU(2)\times SU(2)$, and $J~(\geq 0)$ takes integer or half-integer values. The index $M=(m,\prm)$ denotes the multiplicity of the $(J,J)$ representation and thus $m$ and $\prm$ take values from $-J$ to $J$, respectively. The normalization is taken to be $\int d\Om_3 Y^*_{J_1 M_1}Y_{J_2 M_2} = \dl_{J_1J_2}\dl_{M_1 M_2}$, where $\dl_{M_1 M_2}=\dl_{m_1 m_2}\dl_{\prm_1 \prm_2}$. The complex conjugate is given by $Y^*_{J M}= \eps_M Y_{J -M}$, where $\eps_M=(-1)^{m-\prm}$.

The $SU(2)\times SU(2)$ Clebsch-Gordan coefficient defined by the volume integral of three products of scalar harmonics over $S^3$ is given by
\begin{equation}
     \C^{JM}_{J_1M_1,J_2M_2}
      = \sq{\fr{(2J_1+1)(2J_2+1)}{2J+1}} C^{Jm}_{J_1m_1,J_2m_2}
                           C^{J\prm}_{J_1\prm_1,J_2\prm_2} ,
        \label{C function}
\end{equation} 
where $C^{Jm}_{J_1 m_1, J_2 m_2}$ is the standard $SU(2)$ Clebsch-Gordan coefficient \cite{vmk}. It satisfies the relations $\C^{JM}_{J_1M_1,J_2M_2}=\C^{JM}_{J_2M_2,J_1M_1}=\C^{J-M}_{J_1-M_1,J_2-M_2}=\eps_{M_2}\C^{J_1 M_1}_{JM,J_2 -M_2}$, $\C^{JM}_{00,JN}=\dl_{MN}$ and the crossing relation
\begin{eqnarray}
    \sum_{J \geq 0} \sum_M \eps_M \C^{J_1M_1}_{J_2M_2, J-M}  \C^{J_3M_3}_{JM,J_4M_4} 
    = \sum_{J \geq 0} \sum_M \eps_M \C^{J_1M_1}_{J_4M_4, J-M} \C^{J_3M_3}_{JM,J_2M_2}  . 
            \label{crossing relations}                        
\end{eqnarray}

The 15 conformal Killing vectors on $R \times S^3$ are given in the following. The vector that generates the time translation is denoted by $\eta^\mu = (1,0,0,0)$. The six Killing vectors on $S^3$ are given by $\zeta_{MN}^\mu=(0,\zeta_{M N}^j)$, with $\zeta^j_{MN} = i (\V3/4) \times \{ Y^*_{1/2 M} \hnb^j Y_{1/2 N} - Y_{1/2 N} \hnb^j Y^*_{1/2 M}  \}$. Here, we use the index without $J$ in the case of the four-vector index of $J=1/2$ that appears in the conformal Killing vectors and the corresponding generators. The four vectors that generate special conformal transformations are given by $\zeta_M^\mu=(\zeta_M^0,\zeta_M^j)$ with $\zeta^0_M = \sq{\V3} e^{i\eta}Y^*_{1/2 M}/2$ and $\zeta^j_M = -i\sq{\V3} e^{i\eta} \hnb^j Y^*_{1/2 M}/2$. Their complex conjugates are also conformal Killing vectors for special conformal transformations.

At last, we give the product expansion formulas for scalar harmonics:
\begin{eqnarray}
    && Y_{\half M}^* Y_{JN} = \fr{1}{\sq{\V3}}
             \left\{ \sum_{N^\pp} \C^{\half M}_{JN,J+\half N^\pp}Y^*_{J+\half N^\pp}
                     + \sum_{N^\pp} \C^{\half M}_{JN,J-\half N^\pp}Y^*_{J-\half N^\pp} \right\},
                    \nonumber \\
   && \hnb^i Y_{\half M}^* \hnb_i Y_{JN} = \fr{1}{\sq{\V3}}
             \biggl\{ -2J\sum_{N^\pp} \C^{\half M}_{JN,J+\half N^\pp}Y^*_{J+\half N^\pp}
             \nonumber \\
   && \qquad\qquad\qquad\qquad\qquad
             +(2J+2) \sum_{N^\pp} \C^{\half M}_{JN,J-\half N^\pp}Y^*_{J-\half N^\pp} \biggr\} . 
      \label{scalar harmonics product expansion}
\end{eqnarray}
These are used to show the transformation laws in Section 4.

\section{Generators for Tensor Fields}
\setcounter{equation}{0}
\noindent

Here we briefly summarize the generator of the conformal algebra for the traceless tensor field derived in \cite{hh}.

The Weyl action is quantized in the radiation$^+$ gauge,\footnote{ 
The space of the residual symmetry in the radiation gauge $\hnb^i h_{ij}= \hnb^i h_{i0}= h_{00}=0$ is slightly bigger than the space generated by the 15 conformal Killing vectors, and hence we further remove the lowest mode of $h_{i0}$ satisfying $(\hnb^j \hnb_j +2)h_{i0}=0$, namely, $e_{1/2(My)}=0$ in the text. We call this choice the ``radiation$^+$" gauge.
} 
and then the traceless tensor field $h_{\mu\nu}$ is expanded in tensor and vector harmonics on $S^3$ with three types of mode operators: $c_{J(Mx)}$, $d_{J(Mx)}$, and $e_{J(My)}$. The first two modes belong to the $(J+x, J-x)$ representation of $SU(2) \times SU(2)$ with $J \geq 1$, and the third belongs to the $(J+y, J-y)$ representation with $J \geq 1$ ($e_{1/2(My)}$ is removed by gauge conditions), where $x=\pm 1$ and $y=\pm 1/2$ are the polarization indices for a rank-2 tensor and vector, respectively. The index $M=(m,m^\pp)$ denotes the multiplicity for each representation.

The commutation relations are set as $[c_{J_1 (M_1 x_1)}, c^{\dag}_{J_2 (M_2 x_2)} ] = - [d_{J_1 (M_1 x_1)}, d^{\dag}_{J_2 (M_2 x_2)} ] = \dl_{J_1 J_2} \dl_{M_1 M_2}\dl_{x_1 x_2}$ and $[e_{J_1 (M_1 y_1)}, e^{\dag}_{J_2 (M_2 y_2)} ] = -\dl_{J_1 J_2}\dl_{M_1 M_2}\dl_{y_1 y_2}$, and thus $c_{J(Mx)}$ has the positive metric and $d_{J(Mx)}$ and $e_{J(My)}$ have the negative metric. The Hamiltonian is then given by $H = \sum_{J \geq 1} \{ \sum_{M,x} [ 2J c^{\dag}_{J(Mx)}c_{J(Mx)} -(2J+2)d^{\dag}_{J(Mx)}d_{J(Mx)} ] - \sum_{M,y} (2J+1) e^{\dag}_{J(My)}e_{J(My)} \}$.

The generators of special conformal transformations are given by
\begin{eqnarray}
    Q_M &=& \sum_{J \geq 1}\sum_{M_1,x_1}\sum_{M_2,x_2}
            \E^{\half M}_{J(M_1 x_1), J+\half (M_2 x_2)}
          \Bigl\{ \sq{2J(2J+2)} \eps_{M_1} c^\dag_{J(-M_1 x_1)} c_{J+\half (M_2 x_2)}
                        \nonumber  \\
     && -\sq{(2J+1)(2J+3)} \eps_{M_1} d^\dag_{J(-M_1 x_1)} d_{J+\half (M_2 x_2)}
        + \eps_{M_2} c^\dag_{J+\half (-M_2 x_2)} d_{J (M_1 x_1)} \Bigr\}
                        \nonumber \\
     && + \sum_{J \geq 1}\sum_{M_1,x_1}\sum_{M_2,y_2}
            \H^{\half M}_{J(M_1 x_1); J (M_2 y_2)}
             \Bigl\{ A(J) \eps_{M_1} c^\dag_{J(-M_1 x_1)} e_{J (M_2 y_2)}
                   \nonumber  \\
        &&\qquad\qquad\qquad\qquad\qquad\qquad\qquad
                + B(J) \eps_{M_2} e^\dag_{J(-M_2 y_2)} d_{J (M_1 x_1)} \Bigr\}
                     \nonumber  \\
        && + \sum_{J \geq 1}\sum_{M_1,y_1}\sum_{M_2, y_2}
           \D^{\half M}_{J(M_1 y_1), J+\half (M_2 y_2)}
             C(J) \eps_{M_1} e^\dag_{J(-M_1 y_1)} e_{J+\half (M_2 y_2)} ,
                \label{Q_M for tensor field}
\end{eqnarray}
where 
\begin{eqnarray}
    A(J) &=& \sq{\fr{4J}{(2J-1)(2J+3)}},  \qquad
    B(J) = \sq{\fr{2(2J+2)}{(2J-1)(2J+3)}},
               \nonumber   \\
    C(J) &=& \sq{\fr{(2J-1)(2J+1)(2J+2)(2J+4)}{2J(2J+3)}} .
\end{eqnarray}
The $SU(2)\times SU(2)$ Clebsch-Gordan coefficients defined by the volume integrals of three products of tensor harmonics up to rank 2 are given by
\begin{eqnarray}
   \E^{\half M}_{J(M_1 x_1), J+\half (M_2 x_2)}
       &=& \sq{(2J-1)(J+2)} C^{\half m}_{J+x_1 m_1, J+\half+x_2 m_2} 
              C^{\half \prm}_{J-x_1 \prm_1, J+\half-x_2 \prm_2} ,
                \nonumber \\ 
   \H^{\half M}_{J(M_1 x_1); J(M_2 y_2)}
       &=& - \sq{(2J-1)(2J+3)} 
              C^{\half m}_{J+x_1 m_1, J+y_2 m_2} 
              C^{\half \prm}_{J-x_1 \prm_1, J-y_2 \prm_2} ,
                \nonumber \\
   \D^{\half M}_{J(M_1 y_1), J+\half (M_2 y_2)}
       &=&  \sq{J(2J+3)}  C^{\half m}_{J+y_1 m_1, J+\half+y_2 m_2} 
              C^{\half \prm}_{J-y_1 \prm_1, J+\half-y_2 \prm_2} .
\end{eqnarray}
Here, the type $\E$ is defined by the product of a scalar and two tensor harmonics, the type $\H$ is the product of scalar, tensor, and vector harmonics with a derivative, and the type $\D$ is the product of a scalar and two vector harmonics. In the generator, the coefficients with the four-vector index for scalar harmonics appear. The general expressions of these coefficients are given in \cite{hh}.

In order to construct physical states, we have to prepare creation operators that commute with $Q_M$. From (\ref{Q_M for tensor field}), we find that all creation modes do not commute with $Q_M$, except the lowest creation mode of the tensor field with the positive metric $c^\dag_{1(Mx)}$. Thus the rotation-invariant combination of $c^\dag_{1(Mx)}$ gives the lowest-weight states in the Weyl sector. For other modes, we look for the $Q_M$-invariant creation operators constructed in a bilinear form. Such operators and $c^\dag_{1(Mx)}$ will provide building blocks of physical states for the Weyl sector, which have been constructed and classified in \cite{hamada05}.

\section{Building Blocks for Lower $L$}
\setcounter{equation}{0}
\noindent

Here, we write down the building blocks of physical states with conformal weight $2L$ given in (\ref{building block}) for the lower cases of $L$. For $L=1$, they are given by
\begin{eqnarray}
     S^\dag_{1N} &=& \sq{\fr{2}{3}} ( \hbox{$\sq{2b_1}$}-i\hat{p}) a^\dag_{1N} 
                     - \fr{1}{\sq{2}} \sum_{M_1,M_2} \C^{1N}_{\half M_1, \half M_2} 
                                a^\dag_{\half M_1} a^\dag_{\half M_2} ,
                    \nonumber \\
     {\cal S}^\dag_{00} &=& -\sq{2} ( \hbox{$\sq{2b_1}$}-i\hat{p}) b^\dag_{00} 
                     - \fr{1}{\sq{2}} \sum_M \eps_M a^\dag_{\half -M} a^\dag_{\half M}      
\end{eqnarray}
with conformal weight $2$. The building blocks of $L=2$ with conformal weight $4$ are given by
\begin{eqnarray}
     S^\dag_{2N} &=& \sq{\fr{2}{15}} ( \hbox{$\sq{2b_1}$}-i\hat{p}) a^\dag_{2N} 
                     - \sq{2} \sum_{M_1,M_2} \C^{2N}_{\fr{3}{2} M_1, \half M_2} 
                                a^\dag_{\fr{3}{2} M_1} a^\dag_{\half M_2} 
                     \nonumber \\
                 &&   + \fr{2}{\sq{3}} \sum_{M_1,M_2} \C^{2N}_{1 M_1, 1 M_2} 
                                a^\dag_{1 M_1} a^\dag_{1 M_2} ,
                    \nonumber \\
     {\cal S}^\dag_{1N} &=& -\sq{2} ( \hbox{$\sq{2b_1}$}-i\hat{p}) b^\dag_{1N} 
                     - 4 b^\dag_{00} a^\dag_{1N}
                     - \sq{2} \sum_{M_1,M_2} \C^{1N}_{\fr{3}{2}M_1,\half M_2} 
                            a^\dag_{\fr{3}{2} M_1} a^\dag_{\half M_2}      
                    \nonumber \\
                 && + \fr{2}{\sq{3}} \sum_{M_1,M_2} \C^{1N}_{1 M_1,1 M_2} 
                            a^\dag_{1 M_1} a^\dag_{1 M_2} 
                    + 4 \sum_{M_1,M_2} \C^{1N}_{\half M_1,\half M_2} 
                            b^\dag_{\half M_1} a^\dag_{\half M_2} 
\end{eqnarray}
and the building blocks of $L=3$ with conformal weight $6$ are given by
\begin{eqnarray}
     S^\dag_{3N} &=& \sq{\fr{2}{35}} ( \hbox{$\sq{2b_1}$}-i\hat{p}) a^\dag_{3N} 
                     - \sq{2} \sum_{M_1,M_2} \C^{3N}_{\fr{5}{2} M_1, \half M_2} 
                                a^\dag_{\fr{5}{2} M_1} a^\dag_{\half M_2} 
                     \nonumber \\
                 &&   + 4 \sum_{M_1,M_2} \C^{3N}_{2 M_1, 1 M_2} 
                                a^\dag_{2 M_1} a^\dag_{1 M_2} 
                      - \sq{\fr{15}{2}} \sum_{M_1,M_2} \C^{3N}_{\fr{3}{2} M_1, \fr{3}{2} M_2} 
                                a^\dag_{\fr{3}{2} M_1} a^\dag_{\fr{3}{2} M_2} ,
                    \nonumber \\
     {\cal S}^\dag_{2N} &=& -\sq{2} ( \hbox{$\sq{2b_1}$}-i\hat{p}) b^\dag_{2N} 
                     - 4 \sq{3} b^\dag_{00} a^\dag_{2N}
                     - \sq{2} \sum_{M_1,M_2} \C^{2N}_{\fr{5}{2}M_1,\half M_2} 
                            a^\dag_{\fr{5}{2} M_1} a^\dag_{\half M_2}      
                    \nonumber \\
                 && + 4 \sum_{M_1,M_2} \C^{2N}_{2 M_1,1 M_2} 
                            a^\dag_{2 M_1} a^\dag_{1 M_2} 
                    - \sq{\fr{15}{2}} \sum_{M_1,M_2} \C^{2N}_{\fr{3}{2} M_1,\fr{3}{2} M_2} 
                            a^\dag_{\fr{3}{2} M_1} a^\dag_{\fr{3}{2} M_2} 
                    \nonumber \\
                 &&  + 4 \sq{3} \sum_{M_1,M_2} \C^{2N}_{\fr{3}{2} M_1, \half M_2} 
                            b^\dag_{\fr{3}{2} M_1} a^\dag_{\half M_2} 
                     - 4 \sq{15} \sum_{M_1,M_2} \C^{2N}_{1 M_1,1 M_2} 
                            b^\dag_{1 M_1} a^\dag_{1 M_2} 
                     \nonumber \\
                 &&  + 4 \sq{15} \sum_{M_1,M_2} \C^{2N}_{\half M_1, \fr{3}{2} M_2} 
                            b^\dag_{\half M_1} a^\dag_{\fr{3}{2} M_2}  .
\end{eqnarray}


\end{document}